# It's good to be popular in high school: A look at disparities in STEM AP offerings in Northern California public high schools

David Marasco[1] and Bree Barnett Dreyfuss[2]

[1]Foothill College, Los Altos Hills, CA 94062

[2]Amador Valley High School, Pleasanton, CA 94566

In 2018, in response to the proposed elimination of physics at a predominately Hispanic and socioeconomically disadvantaged (SED) high school, the Northern California/Nevada chapter of the AAPT investigated school demographics and their effect on physics offerings in public high schools in our region. As access was a key issue, the focus was on public, non-charter high schools, which are free to students and do not require winning a lottery for attendance. As reported previously,[1] the data revealed that the percentage of Hispanic students and the percentage of SED students at a high school are highly correlated ($r^2$= 0.60). Additionally, these factors could be used as predictors of a school's physics offerings. To determine if the disparities in course offerings extended through other Advanced Placement (AP) STEM classes the data was further analyzed, revealing that as the popularity of an AP exam drops, so do the relative odds of it being offered, when comparing schools with different demographics. A Northern California public high school student is much more likely to get a strong selection of AP STEM classes if their school serves an affluent, non-Hispanic student majority rather than mostly poor, Hispanic students.

The AP program, managed by the College Board corporation, is widespread in high schools in the United States with 38.4% of all 2020 high school graduates and 32.4% of those in California taking at least one AP exam.[2] It gives students the opportunity to take college-level classes in high school, and good scores on its exams may grant college credit, depending upon the policies of the student's future college. Research has shown that students who participate in AP classes and took the exam at the end of the year tend to earn higher college grades in introductory classes, and graduate college at higher rates.[3] The number of AP exams that a student takes has also been found to be an indicator of college GPAs although it should be noted there is a fee associated with each exam the students register for.[3] However, research into these effects needs to carefully examine confounding variables such as students' family wealth, family educational background, overall educational environment, and motivation. For example, Sadler and Tai found that roughly half of the apparent grade gains predicted by AP exam scores in college introductory biology, chemistry, and physics classes could be accounted for by demographic data and prior academic preparation.[4] Similarly, Dougherty, Mellor, and

Jian observed in Texas data a drop in college graduation rate gains by AP students across all ethnicities once prior preparation and socioeconomic status of both the individual students and their schools were considered.[5] Schneider gives an extensive account of the history of the AP program, from its birth as a joint project between a small handful of elite prep schools and Ivy institutions through its spread through the college admissions ecosystem, and the resulting equity implications.[6] Students that take an AP course and sit for an exam that come from low-income families and qualify for the College Board's fee reduction program still tend to enroll in and graduate from four-year institutions at higher rates than those that do not take AP exams.[7] The College Board's Equity and Access Policy statement encourages educators and schools to take additional steps as needed "to ensure their AP classes reflect the diversity of their student population," including those from low-income families.[8]

**Data Collection**

In addition to restricting the sample to public, non-charter schools, schools with fewer than 200 students were eliminated, as they would not be able to support a wide selection of classes. The sample was restricted to counties in Northern California, defined as those north of the dividing line between Monterey, Kings, Tulare, and Inyo counties and San Luis Obispo, Kern, and San Bernardino counties (Fig. 1). This includes cities such as San Francisco, Oakland, San Jose, and Fresno, but excludes Los Angeles and San Diego.

Using the California Department of Education's School Accountability Report Cards,[9] ethnic and socioeconomic data for each school in the sample were extracted for the 2016-17 school year. Course offerings were determined by looking at the University of California Course List search page,[10] which provided uniform labels for high school classes across the region. The following AP STEM classes were included for each school: Calculus AB, Calculus BC, Biology, Chemistry, Physics-1, Physics-2, Physics-C Mechanics, and Physics-C E&M.

**Results & Discussion**

To investigate the disparity in offerings, schools that are <50% Hispanic and <50% SED (cohort 1) were compared to those that are >50% Hispanic and >50% SED (cohort 2). Schools in these two cohorts account for 79% of the overall sample. Previous work has found that school size has an effect on course offerings.[1,11,12] There was effectively no correlation between school size and the demographic categories of interest, with an $r^2$ of 0.033 for number of students and %SED and 0.005 for number of students and %Hispanic. This suggests that the differences seen are not school size artifacts. Previous work has also explored the effect of socioeconomic status and ethnicity on AP STEM offerings.[13,14] The course-offering data are presented in Table I.

| Course | Number of AP Tests taken in 2018 [N] | Cohort 1<br><br><50% Hispanic and <50% SED [n=183] | Cohort 2<br><br>>50% Hispanic and >50% SED [n=141] | Parity<br><br>Cohort 2 % divided by Cohort 1% |
|---|---|---|---|---|
| Calculus AB | 308,538 | 95.1% [174] | 92.9% [131] | 98% |
| Calculus BC | 139,376 | 72.1% [132] | 36.2% [51] | 50% |
| Biology | 259,663 | 83.1% [152] | 72.3% [102] | 87% |
| Chemistry | 161,852 | 68.3% [125] | 48.9% [69] | 72% |
| Physics 1 | 170,653 | 58.5% [107] | 37.6% [53] | 64% |
| Physics 2 | 25,741 | 25.1% [46] | 1.4% [2] | 6% |
| Physics C - Mech | 57,399 | 23.5% [43] | 7.1% [10] | 30% |
| Physics C - E&M | 25,074 | 9.3% [17] | 2.8% [4] | 30% |
| Physics (total) | 278,867 | 73.8% [135] | 38.3% [54] | 52% |
| BC + Bio, Chem & Physics | | 51.4% [94] | 12.8% [18] | 25% |

*Table I - Percentage of schools offering AP classes in each STEM topic, comparing Cohort 1 and Cohort 2. Parity is the ratio of the cohort columns, expressed as a percentage.*

The AP Calculus exams are useful for examining baselines as they do not require the expenses associated with setting up or maintaining science lab classrooms. Also, while at some schools the only person available to teach a science class might be from another science field and therefore hesitant to teach an AP class (e.g. a person with a Biology background teaching AP Physics), this is less of an issue in math. The Calculus AB exam is very popular with over 300,000 participants in 2018, and it is ubiquitous across our sample. The two different populations enjoyed virtual parity simply because it was offered at nearly every high school in the region. With roughly 140,000 exams, Calculus BC was the smallest of the exams in terms of participants, barring the non-Physics-1 physics offerings. While policies differ from school to school, many students take BC as a "second year" class after a year of AB. Whereas 72.1% of cohort 1 offered BC courses, only 36.2% of cohort 2 had this as an option for their students, for a parity (relative odds of offering) value of 50%.

The historical high school science progression has been Biology followed by Chemistry, then Physics. AP Biology is the most-popular lab science AP exam with almost 260,000 students in 2018. Like Calculus AB, AP Biology is offered at a high percentage of schools across both cohorts. This leads to a small parity gap of 87%. Whatever barriers there are to commit to AP lab classes, Biology appears to have done a good job in solving them. The other AP classes are less popular than Biology, with 161,000 taking the Chemistry exam and 171,000 taking Physics-1. In cohort 1, Calculus BC is offered at more high schools than either AP Chemistry or Physics-1, while in cohort 2 those exams are offered at fewer than 50% of the schools. Parity between the cohorts has dropped by this time in the sequence, Chemistry to 72%, and Physics-1 to 64%. The situation is even worse when physics classes beyond Physics-1 are considered.

Physics-2, Physics-C (Mechanics), and Physics-C (E&M) are typically second-year AP classes. Physics-2 is typically taught after Physics-1, and the Physics-C exams require the use of calculus and may require prerequisite physics courses. Table 1 details the reduced demand for these courses, resulting in a low level of offerings in cohort 2. Only a handful of these schools offer Physics-C (Mechanics), Physics-2, and Physics-C (E&M), resulting in wide parity gaps of 30% and lower. As Physics-1 and Physics-2 are recently implemented courses (2014), it is possible that cohort 2 schools are investing in strengthening their Physics-1 programs before rolling out Physics-2. When offerings across the four different AP physics exams are totalled, parity has dropped to 52%, roughly the same as Calculus BC.

Course offering parity between the cohorts declines with the popularity of the exams, from near parity for Calculus AB down to 50% and below for second-year AP courses. A final comparison examines schools that offer Calculus BC, AP Biology, AP Chemistry, and at least one AP Physics. Roughly one-in-two of the <50% Hispanic, <50% SED schools offer this combination, contrasted with only about one-in-eight of the >50% Hispanic, >50% SED schools, a strikingly inequitable result correlated to demographics.

Given the structural inequities revealed by these data, what can we as educators do to address these issues? First, it is important to recognize the existence of cultural taxation, where faculty of color and other employees from minoritized populations are asked to bear the brunt of the work on diversity and equity issues, performing unpaid labor that is not part of their job description.[15] It is therefore also important for teachers and staff with more privileged backgrounds to advocate for systemic changes within their schools. Teachers and staff at schools with underserved populations that do not currently offer AP Physics should work to establish a program as a matter of equal access. At schools with underserved populations with existing AP Physics programs, the students in those classes should ideally reflect the overall school population. While class specific recruitment and directed advertising may seem unpalatable to some, it may be necessary to counteract systemic barriers to AP Physics that prevent these classes from representing the school's population as a whole. Collaboration between teachers of prior courses and AP STEM teachers can help identify students that may not have previously considered an AP course. Finally, in cases where obstacles for implementation cannot be overcome, such as currently not having enough students to offer second year AP physics classes, cooperation with local two-year colleges through dual enrollment should be aggressively pursued. All of the efforts detailed above should be supported by district and school level leadership with funding for both lab equipment and professional development for faculty. It is the responsibility of the district and school level leadership to study and understand how demographic differences across the district may play a role in course offerings, and to take action when inequities are found. Recognizing systemic inequality in physics is but the first step bringing about change.

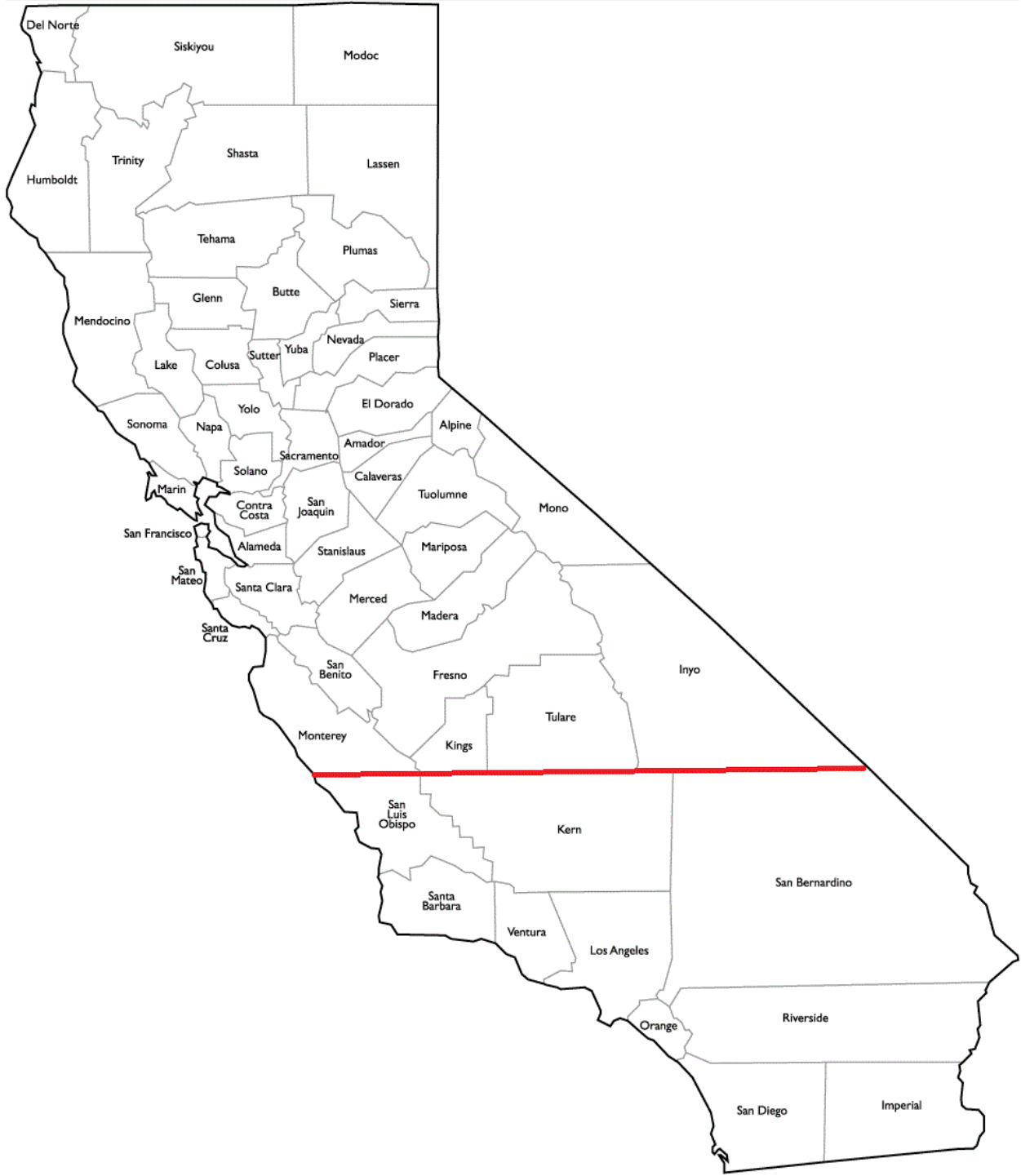

*Fig. 1. Map showing the geographical boundaries of the sample.*